# Generation of sub-40 fs pulses at 1.8 µm by chirp-assisted Raman scattering in hydrogen-filled hollow-core fibre

Sébastien Loranger[1], Philip St.J. Russell[1,2] and David Novoa[1]

[1]Max-Planck Institute for the Science of Light and [2]Department of Physics, Friedrich-Alexander-Universität, Staudtstrasse 2, 91058 Erlangen, Germany

**The possibility of performing time-resolved spectroscopic studies in the molecular fingerprinting region or extending the cut-off wavelength of high-harmonic generation has recently boosted the development of efficient mid-infrared (mid-IR) ultrafast lasers. In particular, fibre lasers based on active media such as thulium or holmium are a very active area of research since they are robust, compact and can operate at high repetition rates. These systems, however, are still complex, are unable to deliver pulses shorter than 100 fs and are not yet as mature as their near-infrared counterparts. Here we report generation of sub-40 fs pulses at 1.8 µm, with quantum efficiencies of 50% and without need for post-compression, in hydrogen-filled hollow-core photonic crystal fibre pumped by a commercial 300-fs fibre laser at 1030 nm. This is achieved by pressure-tuning the dispersion and avoiding Raman gain suppression by adjusting the chirp of the pump pulses and the proportion of higher order modes launched into the fibre. The system is optimized using a physical model that incorporates the main linear and nonlinear contributions to the optical response. The approach is average power-scalable, permits adjustment of the pulse shape and can potentially allow access to much longer wavelengths.**

## 1. Introduction

Lasers emitting ultrashort pulses in the 2 µm range have attracted much interest over the past decade, as they provide an entry point into the mid-IR (2-5 µm)[1]. For example, high-repetition rate Tm-doped fibre lasers[2] are being used to produce high-flux soft X-rays by high-harmonic generation in gases[3]. These complex systems have a relatively narrow gain bandwidth, so that they cannot yet offer pulse durations below 100 fs, unless external spectral broadening and compression is used[4].

Stimulated Raman scattering (SRS) in gases can be used to red-shift the wavelength of a near-IR pump source, offering an alternative approach to generating 2 µm light[5], for example using gas cells[6-8] or gas-filled capillaries[9]. In these systems, linearly chirped pump pulses have been used to decrease the peak power and reduce spectral broadening caused by self-phase modulation (SPM)[10]. As a consequence, the red-shifted Stokes pulses must be re-compressed to reach transform-limited durations, commonly measured by simple intensity autocorrelation, which does not reveal the complex amplitude of the Stokes pulse shape.

The efficiency of gas-based SRS has been dramatically improved with the advent of hollow core photonic crystal fibre (HC-PCF) guiding light by anti-resonant reflection[11]. Offering low transmission loss, tight field confinement and pressure-tunable dispersion, these fibres are extremely advantageous for boosting nonlinear effects in gases[12,13]. HC-PCFs filled with hydrogen have been shown to be highly efficient for down-shifting laser light by 18 THz (rotational SRS)[14] or 125 THz (vibrational)[15], in the latter case allowing access to the 1.8-2 µm region with commonly available 1 µm pump lasers. Recent studies have reported generation of sub-picosecond Stokes pulses in gas-filled HC-PCF[16,17]. Here we show how, by studying in detail the ultrafast nonlinear dynamics that governs the evolution of chirped and unchirped pulses in the gas-filled fibre, these results can be extended into the femtosecond regime. We pay particular attention to detailed measurements of the phase and amplitude of the carrier-wave below the pulse envelope, which are vital in many applications.

In the experiments reported here, 300 fs pulses at 1030 nm from a commercial Yb-based fibre laser are down-shifted to 1.8 µm by vibrational SRS in a $H_2$-filled single-ring HC-PCF. Because the pump pulse duration is less than the lifetime of the Raman coherence, the system operates in the so-called "transient" SRS regime[18], yielding quantum efficiencies in excess of 50%. Moreover, strong interactions with Raman coherence waves (Cw's), along with nonlinear effects such as SPM, enable temporal compression of the 1.8 µm Stokes pulses to durations of 39 fs (measured using second-harmonic generation frequency-resolved optical gating (SHG-FROG)), without need for post-compression. Using numerical modelling to optimize the conversion dynamics, we unveil the key roles played by pump chirp, intermodal walk-off and higher-order modes in the observed dynamics.

## 2. Raman conversion in hydrogen

In SRS the Raman coherence manifests itself as a wave of synchronized internal molecular motion that is driven by the beat-note between pump and Stokes light, which travels at a phase velocity $\Omega_R c / (\omega_P n_P - \omega_S n_S)$, where $\omega_P$, $\omega_R$ and $\Omega_R$ and the angular frequencies of the pump, the Stokes and the Raman coherence, $n_P$ and $n_S$ the refractive indices of pump and Stokes and $c$ the vacuum velocity of light. The coherence lifetime ranges from ps to ns, depending on the gas pressure[19].

In this work we use linearly polarized pump pulses, which favours vibrational SRS, although rotational SRS turns out to play a significant role in the observed dynamics, as explained below. In the transient regime, the pump, Stokes and Cw's are able to exchange energy coherently, causing the Stokes light to be generated predominantly towards the trailing edge of the pulse[18]. Interestingly, in this regime the SRS gain does not saturate with increasing gas pressure, in contrast to the steady-state regime.

Raman gain suppression, normally associated with systems pumped by narrowband lasers, arises when the Cw for pump-to-Stokes generation is identical to the Cw for pump-to-anti-Stokes generation[20,21], i.e., when:

$$\Delta\beta = (\beta_{AS}^{01} - \beta_P^{01}) - (\beta_P^{01} - \beta_S^{01}) = 0, \quad (1)$$

where $\beta_{AS}^{01}$ and $\beta_S^{01}$ and $\beta_P^{01}$ are the wavevectors of the $LP_{01}$-like (from this point on we shall abbreviate "$LP_{mn}$-like" to "$LP_{mn}$") mode at the anti-Stokes, Stokes and pump frequencies. Under these circumstances the probabilities of pump-to-Stokes phonon creation and pump-to-anti-Stokes phonon annihilation are equal, bringing the intramodal Raman gain to zero irrespective of pump power, and favouring the emergence of intermodal SRS, which normally sees much lower gain.

We report that 1.8 μm light can be efficiently generated in the $LP_{01}$ mode by launching a small proportion of pump light into higher-order modes (HOMs) and linearly chirping the pump pulses. The chirp reduces extreme nonlinear spectral broadening, while group velocity walk-off inhibits intermodal SRS, and together they frustrate gain suppression.

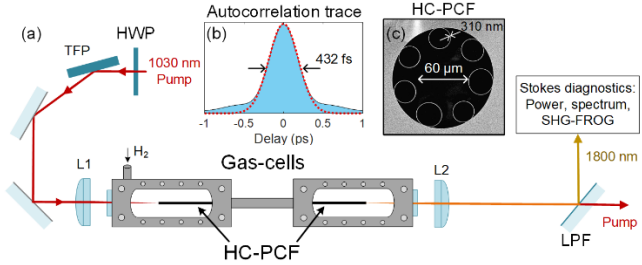

Fig. 1: (a) Experimental set-up. The linear chirp and duration of the pump pulses were precisely controlled using a stretcher inside the fibre laser. A combination of half-wave plate (HWP) and thin-film polarizer (TFP) were used to adjust the input power. Achromatic lenses L1 and L2 were used to launch and collimate the light. A long-pass filter (LPF) with a cut-on wavelength of 1.2 μm separated the Stokes pulses from the residual pump light. (b) Autocorrelation trace of transform-limited pulses. The dashed red curve is a FWHM ~310 fs Gaussian fit. (c) Scanning electron micrograph of the fibre cross-section.

In absence of loss-inducing anti-crossings between the core mode and resonances in the glass walls surrounding the core[22], the modal wavevectors in single-ring HC-PCF are given to good accuracy by the expression[13]:

$$\beta(\lambda) = (2\pi/\lambda)\sqrt{n_{gas}^2(p,\lambda) - \lambda^2 u_{mn}^2/(\pi D)^2} \quad (2)$$

where λ is the vacuum wavelength, $n_{gas}$ is the refractive index of the filling gas at pressure $p$, $u_{mn}$ is the $n$-th zero of a $J_m$ Bessel function and $D$ is the area-adjusted diameter of the core[23]. For the HC-PCF used, calculations based on analytical[20] and numerical modelling (refined to include the effect of anti-crossings[24-26]) predicted that Eq.(1) would be satisfied at a pressure of ~34 bar.

A further attractive feature of the system is that the second vibrational Stokes band at ~7 μm is effectively suppressed by loss of guidance, high glass absorption and low Raman gain, permitting highly efficient conversion solely to the first Stokes band.

## 3. Results & discussion

### A. Experimental results

The single-ring HC-PCF used had core diameter 60 μm and capillary wall thickness 310±40 nm (see inset in Fig. 1). By finite-element modelling we verified that the first anti-crossing occurs at ~650 nm, i.e., sufficiently far away from the first anti-Stokes band ($\lambda_{AS}$ ~ 721 nm) to avoid high loss, while close enough to distort the shallow dispersion and hence to affect the gain suppression condition in Eq. (1). The fibre was pumped by a chirp-adjustable high-repetition rate fibre laser (for details see *Methods*). A low-pass filter was used to separate the Stokes pulses spectrally from the remaining pump light, before being characterized by SHG-FROG and imaged spatially using a thermal camera. A thermal power meter measured the signal power and the spectrum was recorded using an optical spectrum analyser in the visible/near-infrared and an InGaAs spectrometer for longer wavelengths.

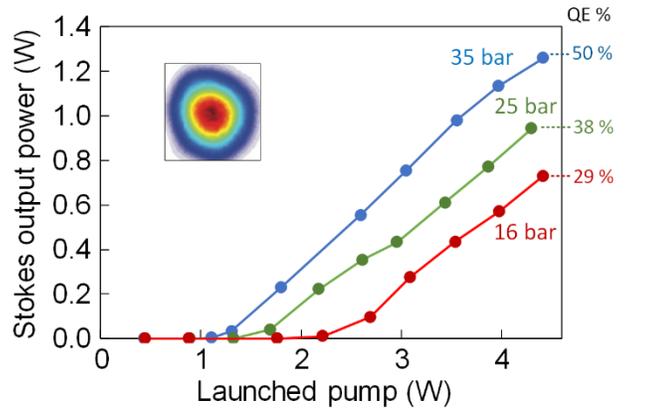

Fig. 2: Mid-IR Stokes power (1.8 μm) as a function of launched pump power (1030 nm) at three different pressures. The laser pulses (300 fs when transform-limited) were chirped to a duration of ~600 fs by adding 0.055 ps² of group-delay-dispersion (GDD). The inset is a near-field image of the generated $LP_{01}$ mode. The quantum efficiencies (%) of Stokes conversion are marked on the right next to each curve.

In the experiment, the highest efficiency of conversion to the $LP_{01}$ Stokes signal was obtained when the pump pulse duration (300 fs when transform-limited) was frequency-chirped to 600 fs by adding 0.055 ps² of GDD (see Fig. 2). As expected in transient SRS[18], the quantum efficiencies (QE) increased with pressure, reaching 50% (>1.2 W) at 35 bar. While the SRS threshold decreased with increasing pressure, the slope efficiency remained unaltered at ~40%. These results were obtained at 151 kHz repetition rate, and the performance of the system remained steady up to ~1 MHz. At higher repetition rates the output power became unstable and ultimately dropped as a consequence of the increasing thermal load.

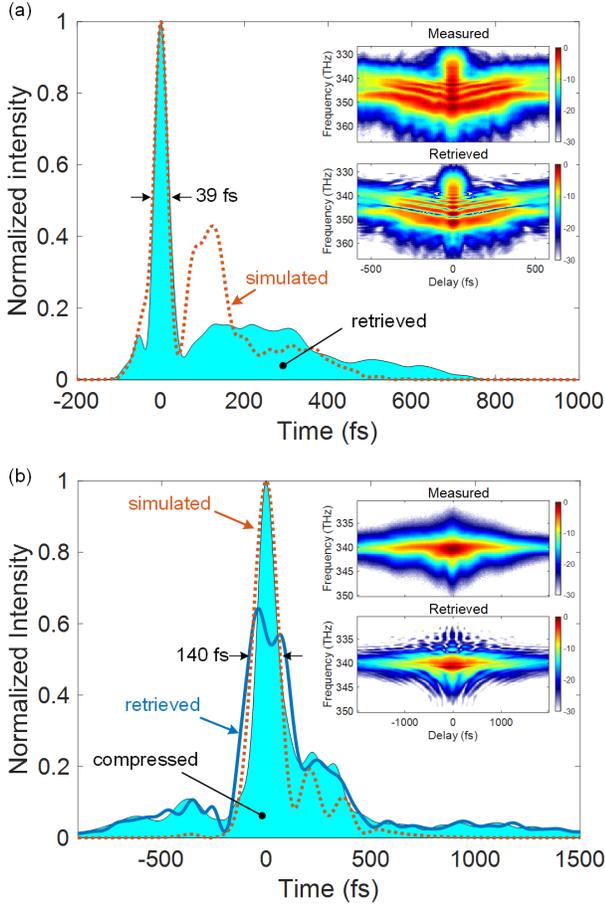

Fig. 3: Stokes pulses generated by pump pulses chirped to durations of (a) ~600 fs (0.055 ps$^2$ of added GDD) and (b) ~1 ps (0.105 ps$^2$ of added GDD). In (a) the shortest feature in the experimental retrieved pulse (under-shaded) is already transform-limited because the output gas-cell window and collimating lens provide the required small negative chirp. In (b) the experimentally retrieved pulse is shown with a solid dark blue line, yielding after compression the under-shaded curve. The under-shaded pulses are compared with simulations (dotted red lines) compressed by adding GDDs of (a) –400 fs$^2$ and (b) –0.014 ps$^2$. Note that the pump chirp is the same in both experiment and simulations.

These results make clear that SRS conversion in gas-filled HC-PCF can be highly efficient, even in the ultrafast regime. There is, however, a surprise hidden in the temporal dynamics. Fig. 3(a) shows the experimental 1.8 μm Stokes pulse retrieved using SHG-FROG when the pump pulse (chirped to 600 fs) energy was 23 μJ. The 1.8 μm pulses unexpectedly show a transform-limited feature, with a duration of 39 fs (full-width half-maximum) and containing 40% of the total energy, followed by a long pedestal.

Figure 3(b) shows the result when the pump pulse is further chirped, to a duration of 1 ps. In this case the short feature in the generated Stokes pulse is 140 fs long.

We found in the experiment that the conversion efficiency increased by a factor of ~2 when the in-coupling was slightly misaligned from optimal LP$_{01}$ injection so as to increase the HOM content (see below).

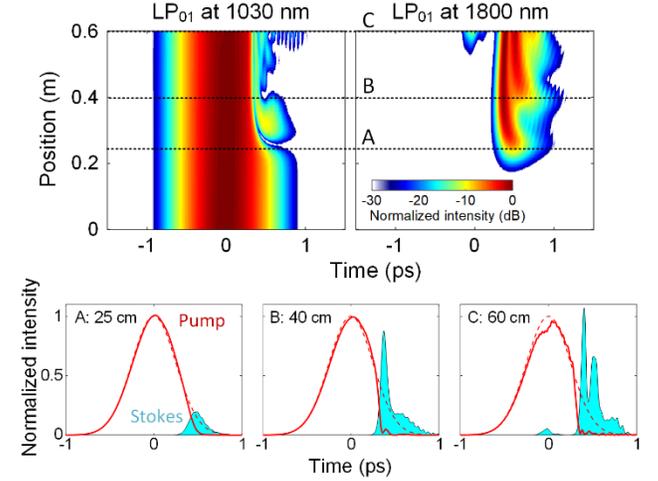

Fig. 4: (a) Simulated dynamics of LP$_{01}$ signals at the pump and Stokes frequencies. The 600 fs chirped pump pulse (top-left panel) contains 28 μJ energy with modal content 80% LP$_{01}$ and 20% LP$_{02}$. (b) Temporal profiles of the pump pulses (upper) and Stokes pulses (lower), multiplied by 2 for clarity, at (A) 25 cm, (B) 40 cm and (C) 60 cm, indicated in the upper plots. The dashed blue curves in the upper plots show the launched pump pulse at $z = 0$ cm.

### B. Numerical modelling

To gain insight into the dynamics of simultaneous Stokes generation and temporal compression we numerically modelled the system using a multimode unidirectional full-field propagation equation[27] (for details see *Methods*). Although SRS plays a major role, instantaneous nonlinear effects such as SPM, combined with the anomalous mid-IR dispersion of the fibre, contribute to self-compression of the Stokes pulses after they have been generated. As a result it was found necessary to include all nonlinear effects, both instantaneous and delayed, for the dynamics to be correctly described. The best agreement with experiment was found when the pump consisted of a mixture of LP$_{01}$ and LP$_{02}$ modes, confirming that imperfect launching conditions play a key role in efficient generation of Stokes light.

For pump pulses chirped to ~600 fs, the ultrashort feature in the Stokes signal (Fig. 3(a)) matches very well in simulations and experiment, showing a transform-limited feature ~39 fs in duration (note that the output gas cell window and collimating lens provided the small GDD required to reach the minimum duration).

When the pump pulse was chirped to a duration of ~1 ps, pump, Stokes and Cw's exchange energy coherently, resulting in a series of Stokes peaks on the trailing edge of the pulse, as seen in Fig 3(b), where dispersion-compensated versions of experimental (under-shaded) and simulated (red dotted line) pulses are compared. The GDD required to compress these pulses is $\sim -0.01$ ps$^2$, which is easily achievable with standard grating compressors. Note that the duration of the transform-limited ultrashort feature in Fig. 3(b) is longer ($\sim$140 fs) than in Fig. 3(a) as a result of the lower peak intensity and weaker nonlinear response. In any case, the 1.8 µm pulse is always shorter than the pump pulse, even immediately after the fibre. As a result its shape can be widely tuned by varying the pressure and the chirp, in response to requirements.

Fig. 4 shows the simulated behaviour of both pump and Stokes signals for pump pulses with 28 µJ pump energy, positively chirped to $\sim$600 fs. The transient behaviour is clear: the pump is depleted on its trailing edge, giving rise to a delayed Stokes pulse. Moreover, since the initial pump duration is longer than a half-cycle at $\sim$125 THz, vibrational coherence is not impulsively excited but takes some time to build up. Once sufficient coherence is present, the Stokes signal is rapidly amplified at the trailing edge of the pump pulse. Combined with the strong remaining pump power, this creates an ultrashort feature on the leading edge of the Stokes pulse, followed by a pedestal, as seen in the lower panels of Fig. 4. At higher pump energies, further Stokes peaks appear after the main one (panel C in Fig. 4), as also seen in the numerical curve in Fig. 3(b).

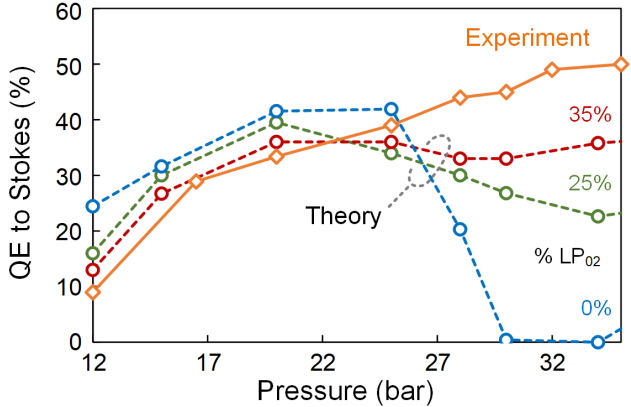

Fig. 5: Measured (diamonds) pressure dependence of the quantum efficiency (QE) of Raman conversion to the 1.8 µm vibrational Stokes band for a 30 µJ pump pulse of chirped duration $\sim$600 fs (0.055 ps$^2$ of GDD added to the laser pulses). The open circles (dashed lines) mark the values calculated numerically for a perfect gas-filled PCF with different proportions of pump energy in the LP$_{02}$ mode.

### C. Coherent gain suppression

As already mentioned, intramodal gain is suppressed when the Cw's for pump-to-Stokes and pump-to-anti-Stokes conversion are identical (blue arrows in Fig. 6(a)). Using narrowband nsec pump pulses[20], the gain suppression pressure was measured to be $\sim$34 bar, in good agreement with the predictions of Eq. (1). This was further confirmed in the femtosecond regime by full-field numerical simulations of the generated Stokes signal with increasing pressure, shown by the blue dashed line in Fig. 5.

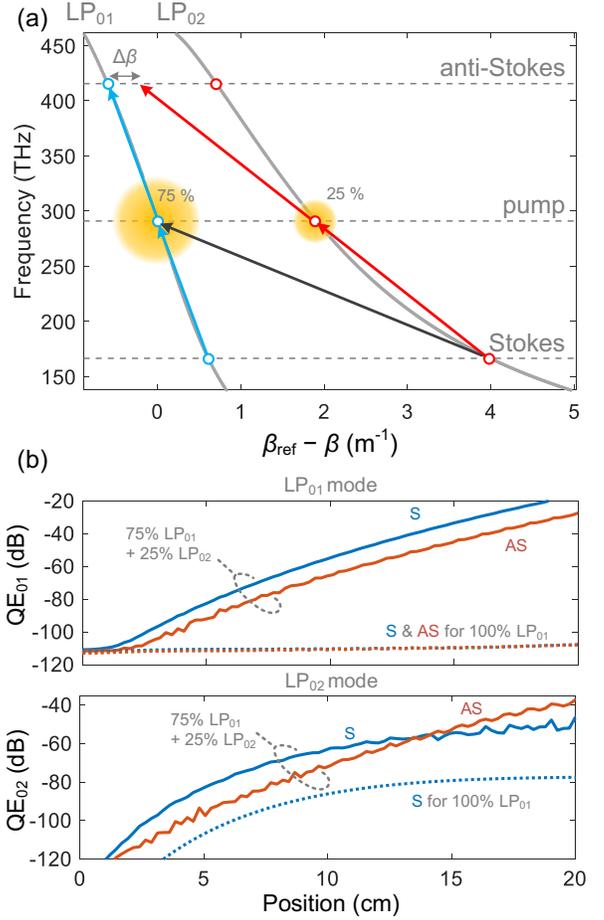

Fig. 6: (a) Dispersion diagram for the LP$_{01}$ and LP$_{02}$ modes at the gain suppression pressure of $\sim$34 bar. To bring out the weak underlying S-shaped dispersion, the quantity $\beta_{ref} - \beta$ is plotted, where $\beta$ is the modal propagation constant and $\beta_{ref} = 1.004\, k_0$. The arrows mark the different Cw's that can be generated by pump-to-Stokes conversion; note that 25% of the pump energy is in the LP$_{02}$ mode (indicated by the shaded circles). Under these conditions some Stokes light is generated in the LP$_{02}$ mode, resulting in a Cw (red) that can be used, over a coherence length of $\pi/\Delta\beta = 8$ mm, to create anti-Stokes photons in the fundamental mode. (b) Simulations of the quantum efficiency for generation of Stokes (S, blue) and anti-Stokes (AS, red) photons in the LP$_{01}$ (upper plot) and LP$_{02}$ (lower plot) modes. Two different cases are compared, the first (full curves) for 25% of the pump energy in the LP$_{02}$ mode, and the second (dashed curves) for 100% of the energy in the LP$_{01}$ mode. In the lower plot the anti-Stokes signal is weaker than –120 dB.

In the experiments, however, we observed no noticeable reduction in Stokes conversion at this pressure (full orange line in Fig. 5). Although SPM and four-wave mixing can alter the Stokes/anti-Stokes balance and partially frustrate gain suppression, for best agreement it was found necessary in the modelling to launch a proportion of pump energy in the LP$_{02}$ mode. Since intermodal SRS (normally weaker than its intramodal counterpart) is strongly enhanced close to the gain suppression point[20,28], a significant Stokes signal is generated in the LP$_{02}$ mode by the gray Cw in Fig. 6(a). This Stokes signal then seeds down-conversion of photons from pump light in the LP$_{02}$ mode, creating the lower red Cw arrow in the figure. This Cw is then able to break the Stokes/anti-Stokes balance by scattering pump photons into anti-Stokes photons via intermodal SRS (upper red arrow in Fig. 6(a)) over a dephasing length of $\pi/\Delta\beta = 8$ mm—significantly longer than the effective gain length, meaning that a few mm of propagation is sufficient to upset the fragile balance needed for coherent gain suppression (see Fig. 6(b)).

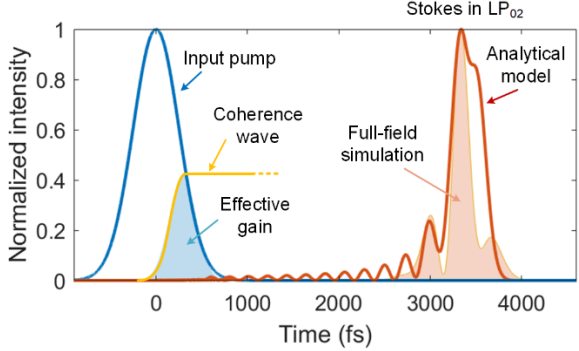

Fig. 7: Pump (LP$_{01}$) and Stokes (LP$_{02}$) pulses after 1 m of propagation. Full-field numerical simulations (Stokes signal is undershaded in light red) are compared with the analytical model in Eq. (3). The simulation was performed using laser pulses carrying 20 µJ of energy entirely in the LP$_{01}$ mode and chirped to a duration of 600 fs by adding 0.055 ps$^2$ of GDD. Both the amplitude of the Raman coherence (which will not decay on the time-scale of the plot) and the effective gain region (blue undershaded) are represented schematically. The gas pressure was 34 bar and intramodal SRS was switched off to reduce pump depletion. Both pump and Stokes pulses are normalized to their peak values.

### D. Effect of pump pulse chirp: Modelling

Remarkably, despite the relatively high LP$_{02}$ mode content used in the simulations in Fig. 5, ~99% of Stokes energy emerges in the LP$_{01}$ mode (Fig. 6(b)). We attribute this to rapid group velocity walk-off of the LP$_{01}$ pump and the LP$_{02}$ Stokes pulses, together with the presence of a linear frequency chirp in the pump pulse that is enhanced by SPM during propagation. During walk-off, the instantaneous beat frequency between chirped LP$_{01}$ pump and LP$_{02}$ Stokes pulses gradually detunes from the Raman frequency, resulting in effective suppression of intermodal Raman gain. This means that a Cw created towards the leading edge of the pump pulse is no longer driven resonantly at the trailing edge by the beat-note between the Stokes and pump pulses.

This effect can be modelled by considering that the intermodal Cw formed by the pump and noise-seeded Stokes light forms an effective gain region at the trailing edge of the pump pulse (blue undershaded in Fig. 7). The growth in the Stokes field $E_s^{02}$ during propagation can then be modelled by integrating over the moving gain region. Assuming that $E_s^{02}$ starts to grow from a very weak signal at $z = 0$, its value at $z = L$ can be written (see Supplementary Material for details):

$$E_s^{02}(t) = \int_{z=0}^{L} g_{\mathit{eff}}\, e^{-(t-t_g-z\Delta v_g/v_g^2)^2/(2\tau_g^2)+i\phi(t-z\Delta v_g/v_g^2,z)}\, dz \quad (3)$$

where $g_{\mathit{eff}}$ is the effective Raman gain, $\Delta v_g$ is the group-velocity difference between the pump and Stokes pulses, $v_g$ is the average group velocity and $\tau_g$ is the temporal width of the effective gain region, offset by time $t_g$ from the pump centre as depicted in Fig. 7. The phase function $\phi(t, z)$ includes the added chirp and SPM induced chirp (see Supplementary Material for details). The time coordinate $t$ is set in the reference frame of the pump pulse. This simple analytical model captures remarkably well the intermodal SRS dynamics of transient pulse generation (see Fig. 7). It also shows that any frequency chirp on pump pulses, for example generated by SPM, will tend to suppress intermodal SRS and lower the efficiency of pump-to-Stokes conversion.

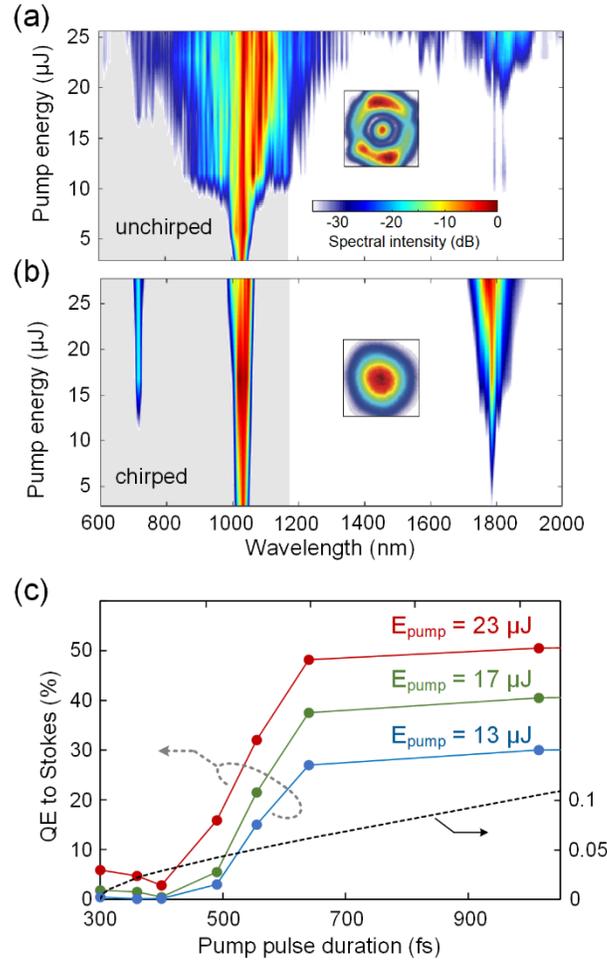

Fig. 8: (a) Measured spectral evolution for different input energies at 34 bar for (upper) transform-limited pump pulses (~300 fs, GDD = 0) and (lower) chirped pump pulses (~600 fs, GDD = 0.05 ps2). The inset shows the far-field beam profile of the signals with wavelengths >~1170 nm. (b) Stokes quantum efficiency measured at 34 bar for different pump pulse energies and input pulse durations, adjusted by adding GDD (right-hand axis) to the 300 fs pulses from the laser. The data points correspond to the total integrated spectral power above 1600 nm. All measurements with efficiencies above 10% (i.e. the detection limit of the thermal camera used) are emitted in a pure fundamental mode.

### E. Effect of pump pulse chirp: Experiment

As seen in the upper panel in Fig. 8(a), transform-limited pump pulses (~300 fs) broaden to a supercontinuum in the fibre and very little 1.8 μm Stokes light is emitted. Although the low conversion efficiency for zero chirp (see Fig. 8(b)) meant that the modal content at 1.8 μm could not be directly measured, at all wavelengths longer than 1200 nm the Stokes light was found to be in the $LP_{02}$ mode (see the far-field profile in the inset of Fig. 8(a), upper). In contrast, when the input pulses were sufficiently chirped, the SRS threshold dropped and the efficiency increased strongly, as seen in Fig. 8(b). Under these circumstances the 1.8 μm signal was always in the $LP_{01}$ mode, while the spectrum consisted of narrowband Stokes and anti-Stokes signals, as seen in the lower panel in Fig. 8(a).

The behaviour for unchirped pump pulses has its origins in the combination of rotational SRS, the reduction in vibrational Raman gain due to pump broadening, and coherent gain suppression. Indeed, in full-field propagation simulations involving both $LP_{01}$ and $LP_{02}$ modes (see *Methods*) the spectral broadening is caused mainly by rotational SRS, seeded by SPM broadening. This is particularly pronounced in the $LP_{02}$ Stokes signal (left column in Fig. (b)) owing to its higher central intensity and stronger dispersion. Under these circumstances the Stokes signal in the $LP_{01}$ mode remains very weak everywhere because (a) it competes with the much stronger intramodal rotational SRS, which is now coherently seeded in the $LP_{02}$-mode, and (b) the Raman gain falls due to pump broadening.

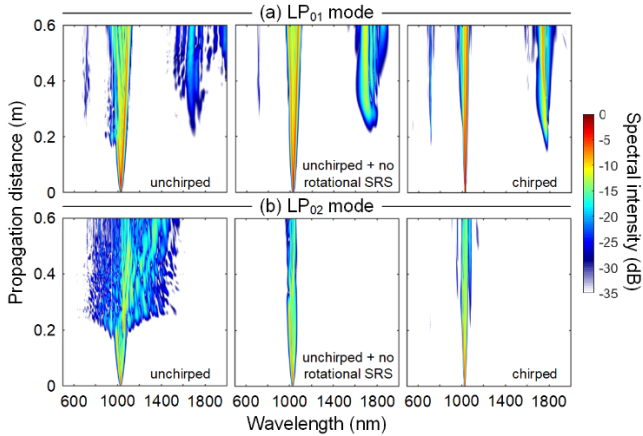

Fig. 9: Simulated evolution of the spectral intensities in (a) the $LP_{01}$ and (b) the $LP_{02}$ modes when 15 μJ pump pulses (75% $LP_{01}$ and 25% $LP_{02}$) are launched into a HC-PCF filled with 34 bar of hydrogen. The left-hand panels are for unchirped ~300 fs pump pulses, including both rotational and vibrational SRS. In the middle panels rotational SRS is switched off, but otherwise the parameters are the same as in (a). In the right-hand panels the pump pulse is chirped to ~600 fs duration, and both rotational and vibrational SRS are included. Simulations using only the $LP_{01}$ mode yield mid-IR Stokes conversion efficiencies << 1% under all conditions.

The key role played by rotational SRS is further confirmed by turning off the rotational Raman gain in the simulations. In this hypothetical case (central column in Fig. 9), the $LP_{02}$ mode is only slightly broadened whereas the $LP_{01}$ Stokes is much more efficiently generated. The situation is radically different when the input pulses are chirped to durations longer than ~600 fs (Fig. 8(b)), when the overall SPM broadening (right-hand column in Fig. 9) is reduced so that the spectrum cannot reach the rotational bands except by noise-seeding, conditions which facilitate vibrational pump-Stokes interactions and favour efficient amplification of the 1.8 μm signal in the $LP_{01}$ mode.

Even in the unrealistic case when rotational SRS is switched off, pumping with a chirped pulse is slightly more efficient (34% quantum efficiency in Fig. 9) than chirp-free pumping (28% quantum efficiency).

## 4. Conclusions

Ultrashort pulses at 1.8 μm can be efficiently generated by pumping hydrogen-filled hollow core PCF with pulses at 1.03 μm from a commercial fibre laser. Quantum efficiencies as high as 50% and 1.8 μm Stokes pulse durations of ~39 fs can be achieved. To understand the underlying physics, ultrafast SRS must be analysed alongside competing nonlinear effects such as SPM, rotational SRS and coherent Raman gain suppression (observed in the ultrafast regime for the first time in this work). All these effects can be kept under control by exciting a proportion of higher order modes in the pump light and chirping the input pulses. We note that coherent gain suppression could potentially be fully eliminated by designing a single-ring HC-PCF so that the first anti-Stokes band lies exactly at an anti-crossing with resonances in the glass walls of the core, causing Eq. (1) to be violated regardless of gas pressure.

The system reported is simple yet robust, making it attractive for generation and spectro-temporal shaping of mid-IR pulses at high repetition rates. Using lasers oscillating at 1.55 μm or longer wavelengths (limited to 2.4 μm for vibrational transitions in hydrogen) it should be possible to generate ultrashort Stokes pulses deep into the mid-IR [29], where optimised HC-PCFs provide excellent mid-IR performance[30,31].

## 5. Methods

### A. Experimental set-up

A 60-cm length of single-ring HC-PCF was mounted between two gas cells to allow evacuation or filling with hydrogen to a controllable pressure (see Fig. 1). The gas cells were equipped with anti-reflection-coated silica windows. The repetition rate of the fibre laser was adjustable and most experiments were conducted at 151 kHz. The pump energy was controlled using a half-wave plate and a thin-film polarizer. The generated signals were collimated, spectrally separated using a low-pass filter with a cut-off wavelength of 1200 nm and delivered to diagnostic systems including an InGaAs spectrometer and a dispersion-free SHG-FROG system for spectro-temporal characterization of the mid-IR pulses. The power and mode profiles were measured with a thermal power meter and a thermal camera. The input chirp of the pump pulses was precisely controlled by fine-tuning the stretcher before the main amplification stage of the fibre laser.

*B. Numerical model*

We modelled the nonlinear dynamics of ultrashort pulses propagating in $H_2$-filled HC-PCF in the transient SRS regime using the multimode full-field unidirectional pulse propagation equation reported in [27]. The model includes full waveguide dispersion, instantaneous third-order nonlinear effects such as Kerr effect, third-harmonic generation, four-wave mixing and the temporally nonlocal contributions of both rotational and vibrational SRS to the nonlinear polarization [32]. Based on experimental observations, only the $LP_{01}$ and $LP_{02}$ modes were included, and the intermodal overlap integrals were calculated using the modal fields obtained from finite-element modelling of a perfect single-ring HC-PCF with the same structural parameters as the fibre used.

## 7. Acknowledgments

We thank Francesco Tani and Daniel Schade for useful discussions and help with various aspects of the project. This work is financially supported by the Max-Planck-Gesellschaft (MPG).

# Supplementary material:
# Generation of sub-40 fs pulses at 1.8 μm by chirp-assisted Raman scattering in hydrogen-filled hollow-core fibre

Sébastien Loranger[1], Philip St.J. Russell[1,2] and David Novoa[1]

[1]Max-Planck Institute for the Science of Light and [2]Department of Physics, Friedrich-Alexander-Universität, Staudtstrasse 2, 91058 Erlangen, Germany

*Analytical model of the effect of chirp and intermodal walk-off on the generation of HOM Stokes pulses*

Let us consider that a Raman coherence wave with a certain profile is formed at the trailing edge of the pump pulse, as shown in Fig. 7 in the manuscript. Assuming negligible pump depletion, a HOM (taken here to be the LP$_{02}$ mode) Stokes pulse will then experience amplification within an effective gain region formed by the overlap of the LP$_{01}$ pump pulse and the coherence wave. This simplified scenario can be modelled by the following differential equation:

$$\frac{\partial E_s^{02}(t,z)}{\partial z} = g_0 Q(t-z\Delta v_g / v_g^2, z) E_p^{01}(t-z\Delta v_g / v_g^2, z) \quad \text{(S1)}$$

where $E_p^{01}$ is the temporal envelope of the pump pulse including chirp, $Q$ is the real-valued temporal profile of the coherence wave envelope, $g_0$ is the Raman gain coefficient, $\Delta v_g = v_g^{01} - v_g^{02}$, $v_g^{mn}$ is the group velocity of the LP$_{mn}$ mode and $v_g$ is the mean of the pump and Stokes group velocities. The time coordinate $t$ is in the time-frame of the pump pulse. Note that Eq. (S1) is only strictly valid in the narrowband limit when the spectra of both pump and Stokes pulses do not significantly overlap, as is the case in our system. The pump pulse is considered to be Gaussian in shape:

$$E_p^{01}(t,z) = \sqrt{P_{pk}} e^{-t^2/(4\tau_0^2 C) + i\phi(t,z)}, \quad \text{(S2)}$$

with peak power $P_{pk}$ and transform-limited duration $\tau_0$, and a phase profile $\phi$ that includes the added GDD $\eta$ and SPM:

$$\phi(t,z) = \eta t^2 / (4\tau_0^4 C) + \gamma \left|E_p^{01}(t,z)\right|^2 z, \quad \text{(S3)}$$

where $C = 1 + \eta^2 / (4\tau_0^4)$, and $\gamma$ is the nonlinear fibre parameter. For simplicity and without loss of generality, we disregard scattering to the second Stokes band and consider that the effective gain region can be approximated by a Gaussian function of width $\tau_g$, offset by time $t_g$ from the pump center:

$$\frac{\partial E_s^{02}(t,z)}{\partial z} = g_{eff} \, e^{-(t-t_g-z\Delta v_g/v_g^2)/(2\tau_g^2)} e^{i\phi(t-z\Delta v_g/v_g^2,z)}, \quad \text{(S4)}$$

where $g_{eff}$ is the effective gain, scaled to account for the strength of both the coherence wave and the pump. The instantaneous deviation from the carrier wave frequency is given by:

$$-\frac{\partial \phi}{\partial t} = -\left( \frac{\eta}{2\tau_0^4 C} + \frac{z\gamma P_{pk}}{\tau_0^2 C} e^{-(t-z\Delta v_g/v_g^2)^2/(2\tau_0^2 C)} \right)\left(t - z\frac{\Delta v_g}{v_g^2}\right) \quad (S5)$$

where the first term within the large bracket accounts for the linear input chirp of the pump (directly adjusted inside the fibre laser in our experiment, see *Methods*) and the second term describes the cumulative $z$-dependent chirp imposed by SPM. We define $z = 0$ as the position where the coherence wave is fully developed, but the Stokes signal is still weak. The Stokes field at point $z = L$ is then obtained by integrating Eq. (S1), yielding Eq. (3) in the manuscript:

$$E_s^{02}(t) = \int_{z=0}^{L} g_{eff}\, e^{-(t-t_g-z\Delta v_g/v_g^2)^2/(2\tau_g^2) + i\phi(t-z\Delta v_g/v_g^2, z)}\, dz. \quad (S6)$$

When SPM is taken into account, no closed-form solution of Eq. (S6) exists, so that the integral must be numerically evaluated, leading to the results in Fig. 7 in the main text.

In the hypothetical case when SPM can be neglected (for example, when the peak power is low), Eq. (S6) can be expressed as:

$$E_s^{02}(t) = \int_{0}^{L} g_{eff}\, e^{-\frac{(t-t_g-z\Delta v_g/v_g^2)^2}{2\tau_g^2}} e^{i\left[\frac{\eta t^2}{4\tau_0^4 C} - \frac{\eta t \Delta v_g}{2\tau_0^4 C v_g^2}z + \frac{\eta t}{4\tau_0^4 C}\frac{z^2 \Delta v_g^2}{v_g^4}\right]} dz. \quad (S7)$$

Neglecting the small higher order phase term in $z^2$, the integral may be evaluated to yield:

$$E_s^{02}(t, L) = \frac{g_{eff} v_g^2 \tau_g \sqrt{\pi}}{\Delta v_g \sqrt{2}} e^{i\left[\frac{\eta t^2}{2\tau_0^4 C} + \delta\omega_g t\right]} e^{-\frac{t^2}{2\tau_s^2}} \times$$

$$\left\{ \mathrm{erf}\left( \frac{\tau_g^{-1} - i\tau_s^{-1}}{\sqrt{2}} t - \frac{t_g + L\Delta v_g/v_g^2}{\tau_g \sqrt{2}} \right) \right. \quad (S8)$$

$$\left. - \mathrm{erf}\left( \frac{\tau_g^{-1} - i\tau_s^{-1}}{\sqrt{2}} t - \frac{t_g}{\tau_g \sqrt{2}} \right) \right\}$$

where $\tau_s = \tau_0^4 2C/(\eta \tau_g)$ and $\delta\omega_g = \eta t_g/(2\tau_0^4 C)$ is the frequency offset of the gain peak with respect to the pump pulse centre.

For chirped pulses, Eq.(S8) predicts the growth of an asymmetric $LP_{02}$ Stokes pulse, in good agreement with the results of full numerical simulations of pulse propagation (see panel (a) in Fig. S1). Conversely, for transform-limited input pulses ($\eta = 0$) the Stokes signal is continuously and efficiently generated under the gain region (see panel (b) in Fig. S1). Fortunately, as our model predicts, any chirp in the pump pulse inhibits the continuous amplification of $LP_{02}$ Stokes light, rendering intramodal SRS conversion very efficient. This in turn justifies why the efficient Stokes pulses measured in our system were always found to emerge in a pure $LP_{01}$ mode.

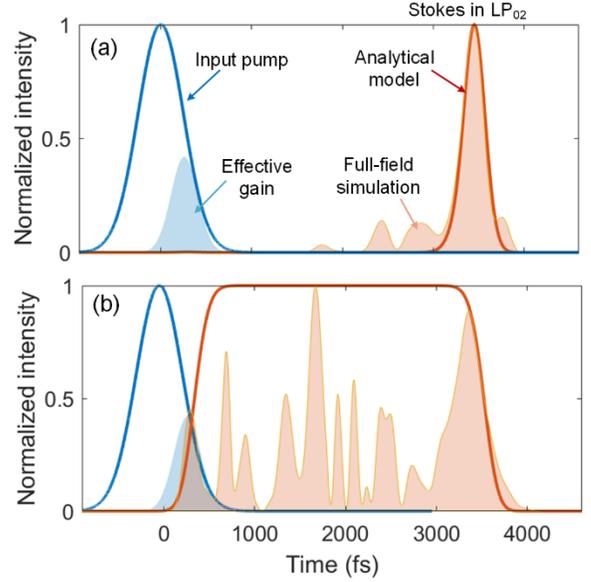

Fig. S1: Comparison between the analytical solution of Eq. (S7) with the full-field simulation for the $LP_{02}$ mode after 1 m propagation excluding rotational, intramodal SRS and the instantaneous Kerr effect. Panel (a) displays the results obtained for a 300 fs transform-limited pump pulse ($\tau_0 = 127$ fs) chirped with $\eta = 0.055$ ps$^2$, yielding a FWHM of 600 fs. Group velocity mismatch is $2.99 \times 10^5$ m/s and average group velocity is $2.99 \times 10^8$ m/s. The gain region was estimated to be 60% of the chirped pump pulse width, hence $\tau_g = 145$ fs. Panel (b) shows the results when a 600 fs transform-limited pump pulse is considered instead ($\tau_0 = 255$ fs, $\eta = 0$). Both pump and Stokes pulses are normalized to their peak values. Note that the actual Stokes signals are very much weaker than the pump, as the analysis only applies to the initial stages of Stokes amplification.